\documentclass[prd,showpacs,preprintnumbers]{revtex4} 
\usepackage{graphicx}
\usepackage{dcolumn}
\usepackage{bm}
\usepackage{latexsym}
\usepackage{amsfonts}
\usepackage{amssymb}
\usepackage{amsmath}
\usepackage{color}
\def\n{d}
\def\p{n}
\begin{document}
\title{
     Kaluza-Klein  braneworld cosmology with static internal dimensions
}
\author{Sugumi Kanno}
\author{David Langlois}
\author{Misao Sasaki}
\author{Jiro Soda}
\affiliation{
 Department of Physics,  McGill University, Montr${\acute e}$al, 
 QC H3A 2T8, Canada
}%
\affiliation{
APC , Universite Paris 7, rue Alice Domon Duquet, 75205 Paris Cedex 13, France
}
\affiliation{
 Yukawa Institute,  Kyoto University, Kyoto, 606-8502
  Japan
}%
\affiliation{
 Department of Physics,  Kyoto University, Kyoto 606-8501, Japan
}%
\date{\today}
\begin{abstract}
  We investigate the Kaluza-Klein braneworld cosmology from the point of view
  of observers on the brane. 
  We first generalize the Shiromizu-Maeda-Sasaki (SMS) equations to
  higher dimensions.  
  As an application, we study a $(4+\p)$-dimensional brane
 with $\p$ dimensions compactified on the brane, in a $(5+\p)$-dimensional  bulk.
  By assuming that  the size of the internal space is static, that the bulk energy-momentum 
  tensor can be neglected, 
  we determine the effect of the bulk geometry on the Kaluza-Klein braneworld.
  Then we derive the effective Friedmann equation on the brane. 
  It turns out that the Friedmann equation explicitly 
  depends on the equation of state, in contrast to the braneworld
 in a 5-dimensional  bulk spacetime. 
 In particular, in a radiation-dominated era,
   the effective Newton constant depends on the scale factor logarithmically.
 If we include a pressureless matter on the brane, this  dependence
 disappears after the radiation-matter equality. This may
 be interpreted as stabilization of the Newton constant by the matter
 on the brane. 
   Our findings imply that the Kaluza-Klein braneworld cosmology is quite
different from the conventional Kaluza-Klein cosmology even at
low energy. 
\end{abstract}
\maketitle
\section{Introduction}
Early universe models are usually motivated by theories trying to describe
 fundamental physics at high energies. String theory is a leading candidate to describe high energy physics
and since it requires 10 dimensions to be consistent, it is natural  to consider 
higher dimensional universes. 
To reconcile a higher dimensional universe
with our empirically 4-dimensional universe,
the traditional approach has been to resort to the Kaluza-Klein compactification.
Another option, the braneworld scenario, was proposed about
a quarter of a century ago~\cite{Akama:1982jy, Rubakov:1983bb}
and has been revived recently, stimulated by the discovery of  D-branes~\cite{Polchinski:1995mt}.
In particular, Randall and Sundrum (RS) proposed an interesting 
framework~\cite{Randall:1999ee,Randall:1999vf}, partly 
inspired by the Horava-Witten model~\cite{Horava:1996ma}. 
RS formulated the universe as a domain wall in 5-dimensional
Anti deSitter (AdS) spacetime. Much work has been done on
its cosmology (see \cite{Langlois:2002bb,Maartens:2003tw,Brax:2004xh,Kanno:2004ns} for reviews) and
black hole
physics~\cite{Chamblin:1999by,Emparan:1999wa,Chamblin:2000ra,Shiromizu:2000pg,
Kudoh:2003xz,Tamaki:2003bq}. However, the RS framework with codimension one braneworld is  insufficient 
to reach 10 dimensions.
To go beyond five dimensions while keeping our spacetime dimensions to four,
we need to consider 
either higher codimensions~\cite{Kanno:2004hr,Charmousis:2005ey}
or Kaluza-Klein compactification on the brane.
The former option, i.e. to realize a higher codimension 
braneworld is difficult due to the strong self-gravity of the brane.
In fact, a higher codimension braneworld develops a severe singularity
except for codimension two models. As a result, no successful
cosmological model is known. Even in the case of codimension two
models, it seems almost impossible to construct a consistent cosmological
model due to the subtlety of the conical
singularity~\cite{Kanno:2004nr,Papantonopoulos:2006dv,
Himmetoglu:2006nw,Kobayashi:2007kv}. 
The latter option, i.e. to consider a Kaluza-Klein cosmology on the
brane~\cite{Charmousis:2004zd,Cuadros-Melgar:2005ex,Papantonopoulos:2006uj,
Yamauchi:2007wm}, is our concern in this paper.

One might think it is a trivial task to construct  braneworld models 
with Kaluza-Klein compactification.
Unfortunately, it is not so~\cite{Chatillon:2006vw}. In the case of the RS
model, the bulk geometry is given and static. Hence, the cosmology on the 
brane is simply due to its motion in the bulk spacetime.
In the case of  Kaluza-Klein braneworlds, however, 
the bulk geometry is not known a 
priori~\cite{Burgess:2001bn,Cline:2003ak,Papantonopoulos:2005nw}. 
Moreover, as we require  the internal space to be static,
we might have to take into account the matter in the bulk, in the form for instance of  fluxes. 
It makes it difficult to solve the bulk geometry in most cases.
In general, we have to solve the bulk geometry and the brane motion 
at the same time and explicit analytical examples are difficult to construct
(see e.g.~\cite{Fabbri:2004mi} for anistropic 5D bulk-brane configurations,  with a problematics 
 similar in spirit to Kaluza-Klein braneworlds). Although  numerical methods seem to be inescapable in general, 
 an analytical approach would be useful even if it is a modest one.
 
In this paper, we make a first step in the analytical description of 
Kaluza-Klein braneworlds. 
Here, we do not intend to solve the bulk geometry.
Instead, we use the Shiromizu-Maeda-Sasaki (SMS) equation~\cite{Shiromizu:1999wj}
to analyze the Kaluza-Klein cosmology.
Of course, this effective equation cannot be solved without knowing
the projected Weyl tensor. Hence, we take the following strategy.
We use the staticity of the internal space as
 a principle to constrain the unknown bulk geometry. We also assume that the bulk matter can be neglected, 
 at least in the vicinity of the branes, in the regimes which we study.  
 Then, we can determine the Friedman equation on the
  brane. Interestingly, the resultant Friedman equation is found to
 depend on the equation of state of the matter explicitly.
 In particular, the effective Newton constant varies logarithmically
at a radiation-dominated stage.
 Thus the Kaluza-Klein braneworld cosmology appears to be
 quite different from the conventional Kaluza-Klein cosmology even
  at low energy. 
  
The organization of this paper is as follows.
In section 2, we consider a braneworld model with the bulk matter
in general dimensions and derive the effective SMS equations on the brane. 
In section 3, we apply the SMS equations to 
 a $(5+n)$-dimensional Kaluza-Klein braneworld model
and derive the effective Friedmann equation on the brane
by imposing the stability of the internal space.
The final section is devoted to conclusion.

\section{SMS effective equation in (\n+1)-dimensions}

In this section, we derive the effective gravitational equations 
on the brane for any dimension. To be as general as possible, 
 we also include a bulk energy momentum tensor.

The action we consider is
\begin{eqnarray}
  S = \frac{1}{2\kappa^2} \int d^{\n+1} x \sqrt{-\tilde{g}}
         \left[ R-2\Lambda\right]
         - \sigma \int d^\n x \sqrt{-g}  + S_m  \,;
\quad \Lambda= -\frac{\n(\n-1)}{2\ell^2} 
\end{eqnarray}
where $\kappa^2$, $\ell$ and $\sigma$ are the gravitational coupling constant,
the scale of the bulk curvature radius and the tension of the brane, respectively.
We assume a negative cosmological constant in the bulk.
Here, $S_m$ represents the action for the matter both 
in the bulk and on the brane. The $(\n+1)$-dimensional and $\n$-dimensional 
metrics are represented by $\tilde{g}$ and $g$, respectively.

We consider  a $\n$-dimensional brane with $(\n-4)$ compactified dimensions.
To describe  the bulk spacetime, we can use Gaussian Normal coordinates so that the 
 metric takes the form
\begin{eqnarray}
  ds^2 = dy^2 + g_{\mu\nu} (y, x^\mu ) dx^\mu dx^\nu \ ,
\end{eqnarray}
and the brane position is  $y=0$ in this coordinate system.
One can deduce the effective equation on the brane
following SMS.
The extrinsic curvature is defined as
\begin{eqnarray}
    K_{\mu\nu} =  - \frac{1}{2}\frac{\partial}{\partial y} g_{\mu\nu}
    \equiv - \frac{1}{2} g_{\mu\nu ,y} \ .
\end{eqnarray}
Using the extrinsic curvature, we can write down the Einstein equations 
in $(\n+1)$-dimensions as 
\begin{eqnarray}
  {}^{(\n+1)}G^y{}_y 
 &=& - \frac{1}{2} R + \frac{1}{2} K^2
 - \frac{1}{2} K^{\alpha\beta} K_{\alpha\beta}
  = \frac{\n(\n-1)}{2\ell^2} + \kappa^2 T^y{}_y   \label{E01}  \,,\\
  {}^{(\n+1)}G^y{}_\mu &=& - \nabla_\lambda K_\mu{}^\lambda{} + \nabla_\mu K
            = \kappa^2 T^y{}_\mu   \label{E02} \,,\\
  {}^{(\n+1)}G^\mu{}_\nu &=& G^\mu{}_\nu
 + \left( K^\mu{}_\nu - \delta^\mu_\nu K \right)_{,y}
  - KK^\mu{}_\nu + \frac{1}{2}\delta^\mu_\nu 
  \left( K^2 + K^{\alpha\beta} K_{\alpha\beta} \right)
  = \frac{\n(\n-1)}{2\ell^2} \delta^\mu_\nu + \kappa^2 S^\mu{}_\nu \delta(y)
  +\kappa^2 T^\mu{}_\nu      \label{E03}          \ ,
\end{eqnarray}
where $G_{\mu\nu}$ is the $\n$-dimensional Einstein tensor,
and $T_{\mu\nu}$, $T_{y\mu}$, and $T_{yy}$ are the components of the bulk
energy momentum tensor.
Here $\nabla_\mu$ denotes the covariant derivative with respect to the metric
$g_{\mu\nu}$, and $S_{\mu\nu}=-\sigma g_{\mu\nu} + t_{\mu\nu}$ is
the energy momentum tensor on the brane, where $t_{\mu\nu}$ is the energy
momentum tensor of the brane matter other than the tension. 
Then, the junction conditions are given by
\begin{eqnarray}
\left[ K^\mu{}_\nu - \delta^\mu_\nu K \right] \big|_{y=0} 
  = \frac{\kappa^2}{2} \left(-\sigma \delta^\mu_\nu + t^\mu{}_{\nu} \right) \ ,
       \label{JC}
\end{eqnarray}
where we have assumed $Z_2$-symmetry.
Combining Eqs.~(\ref{E01}) with (\ref{E03}), we have
\begin{eqnarray}
 && - \frac{1}{\n-1} \left( R_{\mu\nu} - \frac{1}{\n} g_{\mu\nu} R \right)
 = \frac{1}{\n-1} \left[ K_{\mu\nu,y} - g_{\mu\nu} K_{,y} 
        - K K_{\mu\nu} + 2 K_\mu{}^\lambda K_{\lambda\nu} \right]
\nonumber\\
 &&\qquad\qquad
+\frac{1}{\n(\n-1)} g_{\mu\nu} K^2
 + \frac{1}{\n} g_{\mu\nu} K^{\alpha\beta}K_{\alpha\beta}
     - \frac{1}{\ell^2} g_{\mu\nu} - \frac{\kappa^2}{\n-1} \left( 
     T_{\mu\nu}  - \frac{\n-2}{\n}  g_{\mu\nu} T^y{}_y  \right)\,.
     \label{E04}
\end{eqnarray}
The trace of this equation gives
\begin{eqnarray}
  K_{,y} = K^{\alpha\beta} K_{\alpha\beta} - \frac{\n}{\ell^2} 
  + \kappa^2 \frac{\n-2}{\n-1} T^y{}_y - \frac{\kappa^2}{\n-1}  T^\mu{}_\mu
  \label{E05}
\end{eqnarray}
Also the following components of the Weyl tensor are relevant.
\begin{eqnarray}
C_{y\mu y\nu} &=& - \frac{1}{\n-1} \left( R_{\mu\nu}
 - \frac{1}{\n} g_{\mu\nu} R \right)
  + \frac{\n-2}{\n-1} K_{\mu\nu ,y} - \frac{\n-2}{\n(\n-1)} g_{\mu\nu} K_{,y}
  + \frac{\n-3}{\n-1} K_{\mu}{}^\lambda K_{\lambda\nu}
 \nonumber \\
 && \qquad
  + \frac{1}{\n-1} K K_{\mu\nu}
 + \frac{1}{\n} g_{\mu\nu} K^{\alpha\beta} K_{\alpha\beta}
  - \frac{1}{\n(\n-1)} g_{\mu\nu} K^2  \ .
\end{eqnarray}
The above components of the Weyl tensor can be
 rewritten by using Eqs.~(\ref{E04}) and (\ref{E05}) as
\begin{eqnarray}
  C_{y\mu y\nu}
 &=& K_{\mu\nu ,y} - g_{\mu\nu} K_{,y} + K_\mu{}^\lambda K_{\lambda\nu}
  + g_{\mu\nu} K^{\alpha\beta} K_{\alpha\beta} - \frac{\n-1}{\ell^2} g_{\mu\nu}
\nonumber\\
  &&  \qquad+ \kappa^2 \frac{\n-2}{\n} g_{\mu\nu} T^y{}_y - \frac{\kappa^2}{\n-1}
  \left( T_{\mu\nu} + \frac{\n-2}{\n} g_{\mu\nu} T^\alpha{}_\alpha \right)\,.
\end{eqnarray}
Thus off the brane, using these components of the
Weyl tensor, Eq.~(\ref{E03}) is expressed as
\begin{eqnarray}
  G_{\mu\nu} &=& - C_{y\mu y\nu} - K_\mu{}^\lambda K_{\lambda\nu}
  + K K_{\mu\nu} + \frac{1}{2}g_{\mu\nu} K^{\alpha\beta} K_{\alpha\beta}
  -\frac{1}{2} g_{\mu\nu} K^2 + \frac{(\n-1)(\n-2)}{2\ell^2} g_{\mu\nu}  
\nonumber\\
 &&  + \frac{\n-2}{\n-1} \kappa^2 \left( T_{\mu\nu} 
 - \frac{1}{\n}g_{\mu\nu} T^\alpha{}_\alpha \right)
  + \frac{\n-2}{\n} \kappa^2 T^y{}_y g_{\mu\nu} 
\end{eqnarray}
where we stress that the term $T^\alpha{}_\alpha$ is the trace defined with respect to the $\n$-dimensional metric $g$, and not the full trace defined with respect to $\tilde{g}$.
Eliminating the extrinsic curvature by using the junction conditions (\ref{JC}),
and assuming the
RS type relation
\begin{eqnarray}
  \kappa^2 \sigma = \frac{2(\n-1)}{\ell}  \,, 
\end{eqnarray}
we finally obtain the $\n$-dimensional generalization of the SMS equations,
\begin{eqnarray}
\label{einstein_eff}
  R_{\mu\nu} - \frac{1}{2} g_{\mu\nu} R 
  =  - E_{\mu\nu} + 8\pi G\, t_{\mu\nu} + \kappa^4 \pi_{\mu\nu}
   + \frac{\n-2}{\n-1}\kappa^2 \left( T_{\mu\nu} 
   - \frac{1}{\n} g_{\mu\nu} T^\alpha{}_\alpha \right) 
    + \frac{\n-2}{\n} \kappa^2 T^y{}_y g_{\mu\nu}   \ ,
\end{eqnarray}
where we have defined
the Newton constant in $\n$-dimensions by
\begin{eqnarray}
 8\pi G  = \frac{(\n-2)\kappa^2}{2\ell}    \ ,
\end{eqnarray}
the tensor
\begin{eqnarray}
\pi_{\mu\nu} = \frac{1}{4(\n-1)} t t_{\mu\nu}
        - \frac{1}{4}  t_{\mu}{}^\lambda t_{\lambda\nu}
        + \frac{1}{8} g_{\mu\nu} t^{\alpha\beta} t_{\alpha\beta}
        - \frac{1}{8(\n-1)} g_{\mu\nu} t^2\,,    
\label{quadratic}
\end{eqnarray}
and the projected Weyl tensor
\begin{eqnarray}
E_{\mu\nu} = C_{y\mu y\nu} \big|_{y=0} \ .
\end{eqnarray}
These results are also obtained in~\cite{Padilla:2002tg,Cai:2006pa}.

From the Bianchi identity satisfied by the Einstein tensor, we can deduce a constraint equation 
on the tensors that appear on the right hand side of (\ref{einstein_eff}) 
\begin{eqnarray}
  \nabla^\mu E_{\mu\nu}  = \kappa^4 \nabla^\mu \pi_{\mu\nu} 
       + \frac{\n-2}{\n-1} \kappa^2  \nabla^\mu T_{\mu\nu}
       - \frac{\n-2}{\n(\n-1)} \kappa^2 \nabla_\nu T^\alpha{}_\alpha
       + \frac{\n-2}{\n} \kappa^2 \nabla_\nu T^y{}_y \ , 
\label{constraint}
\end{eqnarray}
where we have assumed the conservation of the energy-momentum tensor for the matter on the brane 
\begin{eqnarray}
  \nabla^\mu t_{\mu\nu} = 0  \ ,  
\label{conservation}
\end{eqnarray}
i.e. we forbid the possibility of energy exchange between the brane and the bulk (as studied e.g. 
in \cite{Langlois:2002ke} in 5D brane cosmology).
We also have the conservation law for the bulk matter, which can be decomposed as
\begin{eqnarray}
  0&=& \partial_y T^y{}_y -K T^y{}_y + K^\mu{}_\nu T^\nu{}_\mu 
                + \nabla_\mu T^\mu{}_y        \ , \\
  0&=& \partial_y T^y{}_\mu -K T^y{}_\mu + \nabla_\nu T^\nu{}_\mu \ .
\end{eqnarray}

Of course, the above equations do not form a closed system, because
we do not know $E_{\mu\nu}$. In other words, without knowing the bulk geometry,
we cannot solve the effective Einstein equations (\ref{einstein_eff}). 
However, we may regard the SMS equation as an  initial value equation.
Namely, once $E_{\mu\nu}$ is given from the SMS equation, we can solve
the $(\n+1)$-dimensional Einstein equations along the $y$-direction 
to obtain the bulk geometry. In this picture, unless we impose
some conditions on the properties of the spacetime, there will be 
too many allowed bulk solutions, and most of the solutions will
be physically meaningless. 

In the next section, we will try to solve the effective Einstein's equations to
study the cosmology of  Kaluza-Klein braneworlds from the SMS equation,
by assuming that the internal dimensions are static and that the bulk
energy-momentum tensor can be neglected on the brane.
Given these conditions, we will show that, similarly to the cosmology of
a codimension 1 brane in an empty 5D bulk \cite{bdl99,bdel99}, the effective Friedmann equations 
can be solved, up to a constant of integration. However, in contrast to the
5D bulk, we will not give a bulk geometry associated with this cosmology.

\section{Kaluza-Klein Braneworld Cosmology}

To realize a braneworld in higher dimensions, it seems natural
to consider a codimension one brane with internal dimensions that are compactified 
{\em \`a la} Kaluza-Klein, in short  a Kaluza-Klein braneworld.
Here, for simplicity, we consider $\p$ internal dimensions compactified on a torus. The brane 
thus represents a $(4+\p)$-dimensional spacetime embedded in 
a $(5+\p)$-dimensional spacetime.
Since we wish to study the cosmology, 
we consider  metrics of the form
\begin{eqnarray}
\label{metric}
  ds^2 = - dt^2 + a^2 (t) \delta_{ij} dx^i dx^j + b^2 (t) \delta_{\alpha\beta}dz^\alpha dz^\beta \ ,
\end{eqnarray}
where the $x^i$ are the three ordinary spatial coordinates and the $z^\alpha$ are the internal coordinates. For simplicity, we assume that there is a single 
scale factor $b$ characterizing the size of the internal dimensions. The scale factor $a$ is the usual 
scale factor for the external space. 

We can imagine two kinds of matter, the matter in the bulk
and on the brane.
The bulk matter is important to get a well-behaved  geometry in the 
bulk~\cite{Giddings:2001yu,Kachru:2003aw}.
However, for simplicity, we suppose here that we can ignore it for the cosmology
on the brane. Hereafter, we will consider only the matter on the brane.
Note a recent work, where a similar analysis was considered for a 6D 
Kaluza-Klein brane embedded in a 7D bulk spacetime, and which 
takes into account a bulk energy-momentum tensor and the possibility
of brane-bulk energy exchange \cite{Mazzanti:2007sg}. 
Because of the symmetries, the energy-momentum tensor is
restricted to be of the following form,
\begin{eqnarray}
\label{t}
  t_{\mu\nu} = \left( \rho , P a^2 \delta_{ij} , Q b^2 \delta_{\alpha\beta}\right) \ ,
\end{eqnarray}
where $\rho$ is the energy density, $P$ the external pressure and $Q$ the internal pressure.
Similarly, the projected Weyl tensor is of the form
\begin{eqnarray}
\label{E}
  E_{\mu\nu} = \left( e , \chi a^2 \delta_{ij} , \xi b^2 \delta_{\alpha\beta}\right) \ .
\end{eqnarray}
Moreover, the traceless property of $E_{\mu\nu}$ implies the relation 
$-e+3\chi + \p \xi=0$.
The components of the quadratic tensor  in the energy-momentum tensor (\ref{quadratic}) are given by 
\begin{eqnarray}
\label{pi}
\pi_{00} &=& 
\frac{1}{8(3+\p)} \left[ (\p+2)\rho^2 - 3\p\left( P-Q \right)^2 \right] \ ,
\\
\pi_{ij} &=& \frac{1}{8(3+\p)}\left[ (\p+2)\rho^2 + 2 \rho \left( 2P + \p Q \right)
     +\p \left( P-Q \right) \left( P-3Q \right) \right] a^2 \delta_{ij}
                                                       \ , 
\\
\pi_{\alpha\beta} &=& \frac{1}{8(3+\p)} \left[  (\p+2) \rho^2 + 6 \rho P +2(\p -1)\rho Q 
              +  3\left(\p( P- Q)^2-2Q^2+2PQ \right)  \right] b^2 \delta_{\alpha\beta}  \ . 
\label{pi3}
\end{eqnarray}
Substituting the metric (\ref{metric}) and the tensors (\ref{t}), (\ref{E}) and (\ref{pi}-\ref{pi3}) in the 
effective Einstein equations (\ref{einstein_eff}), one finds 
\begin{eqnarray}
\label{fried1}
 3 H_a^2 + 3\p H_a H_b+\frac{\p(\p-1)}{2}H_b^2 &=& 8\pi G\rho 
   + \frac{\kappa^4}{8(3+\p)} \left[ (\p+2)\rho^2 - 3\p\left( P-Q \right)^2 \right]
   - e \ ,
\\
  -2 \dot{H}_a -3 H_a^2 -2\p H_a H_b - \p\dot{H}_b -\frac{\p(\p+1)}{2}H_b^2
  &=& 8\pi G P 
  +\frac{\kappa^4}{8(3+\p)}\left[ (\p+2)\rho^2 \right.
  \cr
  && \left.
  + 2 \rho \left( 2P + \p Q \right)
     +\p \left( P-Q \right) \left( P-3Q \right) \right]
     -\chi  \ ,
\\
- 3 \dot{H}_a -6 H_a^2 - 3(\p-1) H_a H_b -(\p-1) \dot{H}_b-\frac{\p(\p-1)}{2}H_b^2&=& 8\pi G Q
      + \frac{\kappa^4}{8(3+\p)} \left[  (\p+2) \rho^2 
      + 6 \rho P +2(\p -1)\rho Q \right.
      \cr
 &&     \left.
              +  3\left(\p( P - Q)^2-2Q^2+2PQ \right)  \right]
      - \xi  \ .
\label{fried3}
\end{eqnarray}
In addition to the above equations, we need
the conservation law for the matter (\ref{conservation}), which is given here by
\begin{eqnarray}
   \dot{\rho} + 3H_a  (\rho + P) + \p H_b (\rho +Q) = 0 \ . 
\end{eqnarray}
 The constraint equation for the projected Weyl tensor (\ref{constraint})
in the absence of the bulk matter  can be written as
\begin{eqnarray}  
  && \dot{e} + 3H_a (e+\chi ) + \p H_b ( e + \xi ) 
  +\frac{3\p}{4(3+\p)}\kappa^4 (P-Q) \left[\dot{P} - \dot{Q}+ H_a (\rho +P )-  H_b (\rho +Q)\right]
      = 0 \ , 
\label{Con3}
\end{eqnarray}
where we have defined the Hubble parameters $H_a = \dot{a}/a$ and $H_b = \dot{b} /b$.
In order to integrate explicitly these equations, we will assume 
simple equations of state for the  anisotropic  fluid, namely $P= w \rho$ and $Q=v \rho$, 
where $w$ and $v$ are constants.
Then the equations (\ref{fried1}-\ref{fried3}) become
\begin{eqnarray}
3 H_a^2 + 3\p H_a H_b+\frac{\p(\p-1)}{2}H_b^2 &=& 8\pi G\rho 
+\frac{\kappa^4}{8(3+\p)}\left\{2+\p(1-3(v-w)^2) \right\}\rho^2 
- e \ ,  
\label{F1} \\
  -2 \dot{H}_a -3 H_a^2 -2\p H_a H_b - \p\dot{H}_b -\frac{\p(\p+1)}{2}H_b^2
  &=&
  8\pi G w\rho 
  +\frac{\kappa^4}{8(3+\p)}\left\{2+4w+
  \right. 
  \cr
  &&
  \left.
  +\p (1+2v+3v^2-4vw+w^2) \right\}  \rho^2 -\chi  \ ,  
\label{F2} \\
- 3 \dot{H}_a -6 H_a^2 - 3(\p-1) H_a H_b -(\p-1) \dot{H}_b-\frac{\p(\p-1)}{2}H_b^2&=&
 8\pi G v \rho 
 +\frac{\kappa^4}{8(3+\p)}\left\{2-2v-6v^2+6w+6vw+\right.
 \cr
 &&
 \left.
 +\p(1+2v+3v^2-6vw+3w^2) \right\} \rho^2 
       -\frac{1}{\p}\left(e - 3\chi\right)  \ , 
\label{F3}
\end{eqnarray}
and the conservation law reduces to
\begin{eqnarray}
   \dot{\rho} + 3 (1+w) H_a  \rho + \p (1+v ) H_b \rho = 0 \ . 
\label{Con1}
\end{eqnarray}
The constraint (\ref{Con3}) is written as
\begin{eqnarray}
&&   \dot{e} + 3H_a (e+\chi ) + H_b ( (\p+1)e-3\chi )
  + \frac{3\p}{4(3+\p)}\kappa^4\left[ (w-v )^2 \rho \dot{\rho} 
  +  (H_a (1+w)-H_b (1+v ))(w-v) \rho^2\right]   = 0 \ . 
\label{Con2}
\end{eqnarray}

Since we do not know $e$, $\chi$ and $\xi$, we cannot solve
the above equations. We need to solve the bulk geometry to obtain
 $e$, $\chi$ and $\xi$ in general.
Here, instead of solving the bulk geometry,
we impose the stability of the internal space, that is, we put $b=1$.
By doing so, we may have a singularity in the bulk.
However, if we allow the existence of  matter in the bulk,
it is reasonable to expect that the bulk geometry can be made regular
by a suitable choice of the bulk matter.
What kind of matter is necessary is a different issue. Here we
assume the staticity of the internal space, and seek for an
effective Friedman equation on the brane.
Under this staticity assumption, Eqs.~(\ref{F1}) - (\ref{Con2}) reduce to
\begin{eqnarray}
&&  3 H_a^2  
= 8\pi G\rho 
+\frac{\kappa^4}{8(3+\p)}\left\{2+\p(1-3(v-w)^2) \right\}\rho^2 
- e\ ,
\label{FN1}\\
&&  -2 \dot{H}_a -3 H_a^2 
  = 
  8\pi G w\rho 
  +\frac{\kappa^4}{8(3+\p)}\left\{2+4w
  +\p (1+2v+3v^2-4vw+w^2) \right\}  \rho^2 -\chi 
    \ , 
\label{FN2}\\
&&  - 3 \dot{H}_a -6 H_a^2 = 
8\pi G v \rho 
 +\frac{\kappa^4}{8(3+\p)}\left\{2-2v-6v^2+6w+6vw
 +\p(1+2v+3(v-w)^2) \right\} \rho^2 
       -\frac{1}{\p}\left(e - 3\chi\right) \ ,
\label{FN3}
\end{eqnarray}
and
\begin{eqnarray}
 && \dot{\rho} +    3 (1+w) H_a  \rho  = 0  \ , 
\label{ConN1}\\
 &&     \dot{e} + 3H_a (e+\chi ) 
  + \frac{3\p}{4(3+\p)}\kappa^4\left[ (w-v )^2 \rho \dot{\rho} 
  +  H_a (1+w)(w-v) \rho^2\right]   = 0
  \ .
\label{ConN2}
\end{eqnarray}

What we want is the effective Friedman equation
in the Kaluza-Klein braneworld. For that purpose, a glance of Eq.~(\ref{FN1})
tells us that we need to know $e$, which is a component of the
projected Wely tensor which encodes some information about  the bulk geometry.
For general $w$, Eq.~(\ref{ConN1}) is solved to give
\begin{eqnarray}
  \rho=a^{-3(1+w)} \,,
\end{eqnarray}
where we have absorbed a constant factor into the scale factor by
rescaling it. This is a standard result.
Eliminating $\dot{H}$ and $H^2$ from Eqs.~(\ref{FN1})-(\ref{FN3}), we obtain
\begin{eqnarray}
  -\p\xi = 3\chi -e
     =  8 \pi G \frac{\p}{2+\p} (3w-2v-1) \rho 
  + \kappa^4\frac{\p}{4(3+\p)} v\left(1- 3 w + 3 v  \right) \rho^2\,. 
\label{sigma}
\end{eqnarray}
Thus the components of the projected Weyl tensor 
are related to  the matter on the brane.
Substitution of the above result (\ref{sigma}) into
the constraint equation for the dark radiation (\ref{ConN2}) gives
the equation
\begin{eqnarray}
 \dot{e} + 4 H_a e = -8 \pi G \frac{\p}{2+\p} (3w-2v-1)H_a  \rho
 - \kappa^4\frac{\p}{4(3+\p)} \left( 1- 3 w + 3 v \right)
              \left\{ v +3 (1+w) (w-v) \right\} H_a \rho^2\,.
\end{eqnarray}
This  equation can be integrated easily.
Note that the cases $w=1/3$ and $w=-1/3$ need to be considered separately.

First, we consider the generic case $w\neq 1/3$, $-1/3$.
The solution is given by
\begin{eqnarray}
  e &=& -\frac{3C}{a^4} 
  + 8 \pi G \frac{\p}{2+\p} \frac{1+2v-3w}{1-3w}  a^{-3(1+w)} 
\nonumber\\
 && \quad + \kappa^4\frac{\p}{8(3+\p)} \frac{(1 - 3 w + 3 v)}{(1+3w)} 
              \left\{ v +3 (1+w) (w-v) \right\}
                        a^{-6(1+w)}    \ , 
\label{darkfinal}
\end{eqnarray}
where $C$ is a constant of integration which can be interpreted
 as  dark radiation~\cite{bdel99,Mukohyama:1999qx}. 
Substituting Eq.~(\ref{darkfinal}) into Eq.(\ref{FN1}),
 we obtain the effective Friedman equation
\begin{eqnarray}
    H_a^2 = \frac{8\pi G_{\rm eff}}{3} \rho
    + A \rho^2 + \frac{C}{a^4}  \ ,
\label{effFriedmann}
\end{eqnarray}
where the coefficients are given by
\begin{eqnarray}
  G_{\rm eff}
 &=&  \frac{2(1-3w-\p v)}{(2+\p)(1-3w)} G \ , 
\\
 A&=& \frac{2+6w+\p(1+2v+3(v-w)^2)}{24(3+\p)(1+3w)}\kappa^4 \ .
\end{eqnarray}
The above equations include the well-known 5D case, corresponding to $\p=0$ and for which 
$G_{\rm eff}=G$ and $A=\kappa^4/36$. By contrast, in higher dimensions, the effective Newton constant 
that we have defined depends on the equation of state.  It means that the Kaluza-Klein braneworld
cosmology, provided our assumptions are valid, is different from the conventional Kaluza-Klein cosmology
even at low energies. For $w=0$, we have to assume $nv<1$ in order to have the positive
 effective Newton constant. 
It should be noticed that the effective Newton constant
becomes negative in the regime $0< nv < 1$ and $(1-nv)/3 < w < 1/3$. This indicates some
transient instability around the matter-radiation equality. 

Before proceeding, however, it is important to note that in addition to the
 assumption of staticity of the internal space, we have also assumed that 
the bulk matter can be neglected and that there is no explicit coupling between the matter on the brane and
the matter in the bulk. If we relax one or several of these assumptions, the conclusion will be
altered. 

Now, we consider the radiation-dominated case $w=1/3$. 
In this case, the solution reads 
\begin{eqnarray}
  e = -\frac{3C}{a^4} + {8\pi G}\frac{2\p}{2+\p}v \frac{\log a}{a^{4}} 
   - \kappa^4\frac{\p}{16(3+\p)} v ( 9v-4 )  a^{-8}    \ . 
\label{darkfinal2}
\end{eqnarray}
There appears a logarithmic correction in the above. 
Substituting Eq.~(\ref{darkfinal2}) into Eq.~(\ref{FN1}),
 we obtain the effective Friedman equation (\ref{effFriedmann})
with the coefficients given by
\begin{eqnarray}
 G_{\rm eff} 
 &=&  G\left(1  - \frac{2\p}{2+\p}v \log \frac{a}{a_{*}}\right) \ , \\
 A&=&   \kappa^4 \frac{12+\p(4 + 9 v^2)}{144(3+\p)}  \ ,
\end{eqnarray}
where  $a_{*}$ is a constant of integration corresponding to the dark radiation 
component $C$.
Remarkably, the effective Newton constant depends logarithmically on the scale factor.
This interesting behavior of the cosmological
evolution occurs only during a  radiation-dominated stage.
Furthermore, depending on the value $a_*$, 
the effective Newton constant may become negative.
This implies that the dark radiation component should be
chosen appropriately in order to realize a sensible cosmology on the brane.
It may be mentioned that this behavior is in contrast to the case of
the Brans-Dicke cosmology in which the effective Newton constant
begins to depend on time after pressureless matter starts to dominate.
In a sense, one may say that dark matter stabilizes the effective 
gravitational coupling constant. The relevant constraint
comes from  nucleosynthesis~\cite{Copi:2003xd,Uzan:2004qr}:
\begin{eqnarray}
 \frac{ \Delta G_{\rm eff} }{G_{\rm eff}} = 0.01^{+0.20}_{-0.16}\,,
\end{eqnarray}
at one-sigma confidence level.
It is easy to see that this constraint is 
satisfied for a sufficiently large $a_*$.

Finally, we consider the case $w=-1/3$. 
The solution is given by 
\begin{eqnarray}
  e = -\frac{3C}{a^4} + {8\pi G}\frac{\p}{2+\p}(1+v)a^{-2}
        + {\kappa^4}\frac{\p}{12(3+\p)} ( 2+ 3 v )^2
                        \frac{\log a}{a^{4}}    \ , 
\label{darkfinal3}
\end{eqnarray}
where a logarithmic term appears again.
 Substituting Eq.~(\ref{darkfinal3}) into Eq.~(\ref{FN1}),
 we obtain the coefficients in the effective Friedman equation as
\begin{eqnarray}
 G_{\rm eff} 
 &=&  \frac{2-\p v}{2+\p} G   \ , 
\\
 A&=&  \kappa^4\frac{6 + \p(2 - 6v - 9 v^2)}{72(3+\p)}
       -   {\kappa^4}\frac{\p}{36 (3+\p)}( 2+3v )^2
         \log \frac{a}{a_{*}}   \ ,
\end{eqnarray}
where $a_*$ is the constant of integration we mentioned before. 
In this case, we have a logarithmic scale factor 
dependence in the coefficient of $\rho^2$. 
This can have some impact at high energy. Therefore, its effect on
the inflationary scenario may be interesting. 

Apparently, the Kaluza-Klein cosmology we have obtained
is different from the conventional Kaluza-Klein cosmology.
In particular, the effective Friedman equation depends on 
the equation of state of the matter explicitly. 
This result will  hold in more general
Kaluza-Klein spacetimes. The reason is the following.
The constraint equation (\ref{constraint}) in the absence
of the bulk matter reads
\begin{eqnarray}
    \nabla^\mu E_{\mu\nu} = \kappa^4 \nabla^\mu \pi_{\mu\nu}  \ . 
\label{sol}
\end{eqnarray}
Hence, in general,  the projected Weyl tensor 
is affected by the energy momentum tensor on the brane.   
If the brane was isotropic and homogeneous, the matter part 
would have the additional property that $\nabla^\mu \pi_{\mu\nu}=0$.
 In our example, this can be seen by setting $P=Q$ in Eq.~(\ref{Con3}).
The effect of matter would then not appear in Eq.~(\ref{sol}).
 This is related to Birkoff's theorem.
 Because of  spherical-like symmetry, one does not see the details of the
matter contents, but rather see the dark radiation which is independently 
conserved. 
On the contrary, in the case of  Kaluza-Klein braneworlds
this no longer happens because of the anisotropy of the brane.
In particular, $\pi_{\mu\nu}$ is no longer conserved,
$\nabla^\mu \pi_{\mu\nu} \neq 0$.
Thus the anisotropic brane will deform the bulk geometry
nontrivially. Then the back-reaction of this to the brane 
will modify the effective Friedman equation. Because of
the strong anisotropy of the Kaluza-Klein braneworld,
we then expect that the modification of
the Friedmann equation will persist even at low energies.

Since the SMS equation is not a closed system, 
we cannot formulate cosmological
perturbation theory without further information.
As we noted before, we may regard the SMS equations as giving
the ``initial data'' to solve the bulk geometry. 
Of course, in general, there would be a singularity in the bulk.
However, if one introduces some bulk matter,
one may obtain a non-singular geometry in the bulk.
With some explicit bulk matter, one can also formulate explicitly the cosmological perturbation theory
for the corresponding Kaluza-Klein braneworld.
Since there is the  possibility of a deviation from Newton's
law at low energy, it would be interesting to perform this program explicitly.

\section{Conclusion}

We have considered a $(\n+1)$-dimensional gravitational system with
 bulk matter and a $\n$-dimensional brane,
and derived the effective $\n$-dimensional Einstein equations on the brane.
As an application, we have studied the cosmology of Kaluza-Klein braneworld, 
with $\p$ internal toroidal dimensions. By neglecting the bulk matter and 
imposing the staticity  of the internal space, we have obtained a closed set of equations, from 
which we have been able to  derive  the effective Friedmann equation.
We have found that the resultant Friedmann equation explicitly depends
on the equation of state.
In particular, if the universe is dominated by radiation,
a resonant contribution to the projected Weyl tensor gives a
time variation to the effective gravitational coupling constant.
This time dependence disappears after the radiation-matter equality,
which can be interpreted as a
  stabilization of the Newton constant by the matter on the brane.
It should be emphasized that
the Kaluza-Klein braneworld cosmology is quite
different from the conventional Kaluza-Klein cosmology even at
low energy. It is in contrast to the fact that the conventional 
Friedmann equation can be recovered at low energy in the RS braneworld
cosmology. Hence, it is important to see whether Newtonian gravity
 can be recovered~\cite{Garriga:1999yh} on the Kaluza-Klein braneworld. This question 
 has already been studied in some specific 6D models based on flux compactifications \cite{Peloso:2006cq,Kobayashi:2007kv}.

We have also discussed the method to obtain the bulk geometry from
the brane data.
It seems always possible to adjust $E_{\mu\nu}$ so that the braneworld 
has a static internal space during the cosmic expansion. 
Of course, in general, the geometry of the bulk will be contrived 
and it will perhaps contain singularities.
However, assuming the presence of  matter in the bulk,
there is a chance to have a non-singular bulk.
Admittedly, it is a non-trivial problem to find an appropriate matter
which can stabilize the braneworld without any pathology in the bulk.
An alternative approach is to start from known bulk solutions in which a brane is embedded. This method has been used very recently to study the cosmology of Kaluza-Klein branes, with one internal dimension, in 6D bulk solutions of Einstein-Maxwell equations \cite{Minamitsuji:2007fx,Papantonopoulos:2007fk}.

There are many other issues to be explored. It is interesting to formulate
the quantum creation of the Kaluza-Klein braneworld as is done in RS
models~\cite{Garriga:1999bq,Koyama:2000vs}. It is also important to
understand the low energy description of the Kaluza-Klein 
braneworld~\cite{Langlois:2002hz,Kanno:2002ia,Kanno,Wiseman:2002nn,Shiromizu:2002qr,Kanno:2003xy,Kanno:2005zr}
and the Kaluza-Klein corrections~\cite{Kanno:2003vf,Minamitsuji:2005xs}.
It is intriguing to consider born-again scenario 
in higher dimensions~\cite{Kanno:2002py,Leeper:2005pg}.
We leave these issues for future studies.

\begin{acknowledgements}
This work was supported by JSPS-CNRS Joint Research Program
and JSPS Grant-in-Aid for Scientific Research (B) No.~17340075.
SK was supported by a grant for Research Abroad by the JSPS. 
JS was supported by JSPS Grand-in-Aid for Scientific Research (C)
No.~18540262.
\end{acknowledgements}

\end{document}